\documentclass[a4paper]{article}
\usepackage{graphicx}
\usepackage{times}
\begin{document}


\title{Differential rotation and meridional flow on the lower zero age main sequence: Reynolds stress versus baroclinic flow}

\author{M. K\"uker , G. R\"udiger 
\\[1cm]
Leibniz-Institut f\"ur Astrophysik Potsdam\\
 An der Sternwarte 16\\
14482 Potsdam, Germany
}

\maketitle


\begin{abstract}
  We study the variation of surface differential rotation and meridional flow along the lower part of the zero age main sequence (ZAMS). We first compute a sequence of stellar models with masses from 0.3 to 1.5 solar masses. We then construct mean field models of their outer convection zones and compute differential rotation and meridional flows by solving the Reynolds equation with transport coefficients from the second order correlation approximation. For a fixed rotation period of 2.5 d we find a strong dependence of the surface differential rotation on the effective temperature with weak surface shear for M dwarfs and very large values for F stars. The increase with effective temperature is modest below 6000 K but very steep above 6000 K. The meridional flow shows a similar variation with temperature but the increase with temperature is not quite so steep. Both the surface rotation and the meridional circulation are solar-type over the entire temperature range. We also study the dependence of differential rotation and meridional flow on the rotation period for masses. from 0.3 to 1.1 solar masses. The variation of the differential rotation with period is weak except for very rapid rotation. The meridional flow shows a systematic increase of the flow speed with the rotation rate. Numerical experiments in which either the $\Lambda$ effect is dropped in the Reynolds stress or the baroclinic term in the equation of motion is cancelled show that for effective temperatures below 6000 K the Reynolds stress is the dominant driver of differential rotation. 
\end{abstract}

\clearpage

\section{Introduction}
Differential rotation is a powerful generator of magnetic fields and therefore a key ingredient in most stellar dynamo models. The solar surface differential rotation can be observed directly and has therefore been known for a long time. More recently, helioseismology has revealed the internal rotation and found that the pattern on the surface persists throughout the convection zone and vanishes in a relatively shallow layer below. 
Surface differential rotation has also been inferred by photometry and spectroscopy for a number of main sequence and giant stars and asteroseismology will soon be able to detect the internal differential rotation of giant stars. The growing amount of observational data now allows more systematic studies in order to understand what factors determine the rotation of a star.

As the classical $\alpha \Omega$ dynamo has problems reproducing all details of the solar activity cycle, a modified version, the flux transport dynamo was proposed (Wang \& Sheeley 1991, Choudhuri et al.~1995, Dikpati \& Charbonneau 1999, K\"uker et al.~2001) and has been studied in detail since. In this dynamo the cycle time strongly depends on the large-scale meridional flow. The latter does not make a direct contribution to the generation of the magnetic field but links the generator of the toroidal field component, the differential rotation, with the producer of the poloidal field, the $\alpha$ effect. 

The solar meridional flow is much slower than the rotation and therefore harder to observe. Recent observations found a polewards flow of about 20 m/s at the surface and in the sub-surface layer (Komm et al.~2005). Mass conservation requires a "return flow" directed towards the equator but neither the depth at which it occurs nor its speed are known from observations. However, the return flow is crucial to the flux transport dynamo as it produces the tilt of the wings in the butterfly diagram of solar activity. 

In addition to the role it plays in stellar dynamos, the meridional flow is also a powerful transporter of angular momentum and thus can drive differential rotation. On the other hand, differential rotation will cause meridional flows unless  the rotation period is constant on cylindrical surfaces parallel to the rotation axis. This causes serious problems for any model trying to explain the differential rotation as the result of Reynolds stress alone. The so-called Taylor number puzzle was solved by Kitchatinov \& R\"udiger (1995) by including a baroclininc force caused by the effect of the Coriolis force on the convective heat transport. Their model reproduces the observed solar rotation pattern remarkably well (K\"uker \& Stix 2001).
 
A key question to theory is which stellar properties determine the differential rotation. Several observational studies have found a rather weak relation between differential rotation and rotation period (Hall 1991, Donahue et al.~1996, Messina \& Guinan 2003, Reiners \& Schmitt 2003).    
Barnes et al.~(2005) found a much stronger dependence on effective temperature, namely
\begin{equation}
  \delta \Omega \propto T_{\rm eff}^{8.92\pm0.31},
\end{equation} 
 which was confirmed by Reiners (2006). Collier Cameron (2007) found a slightly smaller exponent  but a very similar power law dependence of the form
 \begin{equation} 
 \delta \Omega = 0.053   \left(\frac{T_{\rm eff} }{5130 \;{\rm K}} \right) ^ {8.6} {\rm rad/d}.    \label{barneslaw}
\end{equation} 
Kitchatinov \& Olemskoy (2010) computed the surface differential rotation for a series of ZAMS models with masses from 0.4 to 1.2 solar masses. They found a monotonous increase with temperature that qualitatively agrees with the observational findings but did not make any quantitative comparison. In this paper we carry out a similar mass sequence and compare the result to the observational findings. 
\section{Model}
Current mean field models show that in a rotating, stratified convection zone the Reynolds stress is not purely diffusive and will generally cause a non-solid rotation. The non-diffusive part is known as the $\Lambda$ effect (R\"udiger 1989). 
differential rotation driven solely by the Reynolds stress is always solar-type and will drive a meridional flow that is directed towards the poles at the surface. This flow counteracts the stress and reduces the differential rotation compared to a state without meridional flow. Moreover it drives the rotation towards a state where the rotation period is constant on cylindrical surfaces (R\"udiger et al.~1998).       

The Coriolis force also causes the convective heat transport to deviate from spherical symmetry. Instead of the strictly radial transport that would occur in a non-rotating star, the heat flux has a small horizontal component that is directed towards the poles. As a consequence, the gradients of density and pressure are no longer aligned and a baroclinic term appears in the equation of motion. This term drives of meridional flows that are directed towards the equator at the stellar surface (K\"uker et al.~2011) and can cause rotation patterns with solar-type surface differential rotation.      
However, the internal rotation will be disc-shaped, i.e.~the rotation period will be a function of $z$ (the distance from the equatorial plane) only. Together with the wrong direction of the surface flow, this rules out the baroclinic term as the main (or sole) driver of the solar differential rotation.

Though not the main driver of differential rotation the baroclinic term is crucial in avoiding the Taylor Proudman state as it reduces the back reaction of the meridional flow on the differential rotation.
 For fast rotation, the part of the equation for the meridional flow is dominated by the baroclinic and centrifugal terms, i.e.
\begin{equation}
  \sin \theta r \frac{\partial \Omega^2}{\partial z} - \frac{g}{c_p r} \frac{\partial S}{\partial \theta} \approx 0, \label{thermal}
\end{equation} 
where $r$ and $\theta$ are the usual spherical polar coordinates, $\Omega$ is the angular velocity, $S$ the entropy, $c_p$ the specific heat capacity at constant pressure, and $g$ gravity. The system is then said to be in "thermal wind equilibrium." 
With the further assumption that the entropy is constant on isocontours of the angular velocity, it is then possible to integrate such an isocontour from a given starting point (Balbus 2009) and reproduce the tilt  of the angular velocity isocontours in the solar convection zone at intermediate depths (Balbus et al.~2009). 

\begin{figure}
  \begin{center}
  \includegraphics[width=5.5cm]{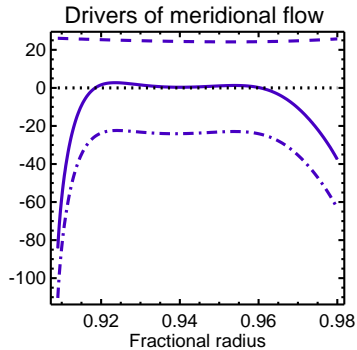}
  \end{center}
  \caption{ \label{wind}
  The two terms driving the meridional flow vs.~radius at $45^\circ$ latitude for a 1.5 $M_\odot$ ZAMS star rotating with a period of 2.5 d. Dashed line: baroclinic term. Dash-dotted line: centrifugal term. Solid line: sum}
\end{figure}

The thermal wind equilibrium does not, however, determine the variation of the angular velocity with latitude or the total shear between the polar caps and the equator. Moreover, Eq.~\ref{thermal} does not hold close to the boundaries of the convection zone.
 Figure \ref{wind} shows the two terms in Eq.~(\ref{thermal}) as functions of the fractional radius of the star at $45^\circ$ latitude for a ZAMS star with 1.5 solar masses. The centrifugal and baroclinic terms cancel each other in the bulk of the convection zone, which means that the Reynolds stress could be neglected there. Close to the boundaries, however, the centrifugal term is larger (by amplitude) than the baroclinic term and the sum of the two terms deviates from zero. As the difference has to be balanced by the Reynolds stress, the deviation indicates its importance. This is a consequence of the stress-free boundary condition which requires that the Reynolds stress, i.e.~the sum of viscosity and $\Lambda$ effect, vanish.  
To compute the rotation of a star we first need a stellar model for which we then compute the rotation and meridional flow patterns. 
We base our model convection zones on models from stellar evolutionary tracks computed with the Mesa/Stars code (Paxton et al.~2011). Our ZAMS models were computed with $Z=0.02$,  $Y=0.28$, and $\alpha_{\rm MLT} = 2$. The model parameters have been chosen to reproduce the luminosity and mass of the present-day Sun at an age of 4.567 Gyr and an outer convection zone that ends at a fractional radius of 0.71.  
Our models cover the mass range from 0.3 to 1.5 solar masses. The 0.3 $M_\odot$ model is fully convective, all other models have outer convection zones. The relative depth of the CZ varies between 56\% for 0.35 $M_\odot$ and 9\% for 1.5 $M_\odot$. The temperature at the bottom of the convection zone decreases from $5.3 \times 10^6$ K at 0.35 $M_\odot$ to $5.0 \times 10^5$ K at 1.5 $M_\odot$. 

We then solve the equations of angular momentum transport, meridional flow, and convective heat transport as described in K\"uker et al (2011) for the model stars we have computed above. The convection zone is approximated by a simple stratification model that assumes adiabaticity and hydrostatic equilibrium in spherical geometry. The Reynolds stress and convective heat transport are computed using standard mixing length theory and the second order correlation approximation (SOCA). We assume axisymmetry and mirror symmetry with respect to the equatorial plane. The mass density depends on radius only but the entropy is a function of radius and latitude. The code solves for a stationary solution for stress-free boundaries and an imposed radial heat flux that is constant with latitude at the lower boundary. The transport coefficients are computed from the stratification and depend on the rotation rate of the star via the Coriolis number,
$
   \Omega^* = {4 \pi}/{\rm Ro}, 
$
where Ro the Rossby number.  
%
%
\section{Results}
\begin{figure}
\includegraphics[width=8.0cm]{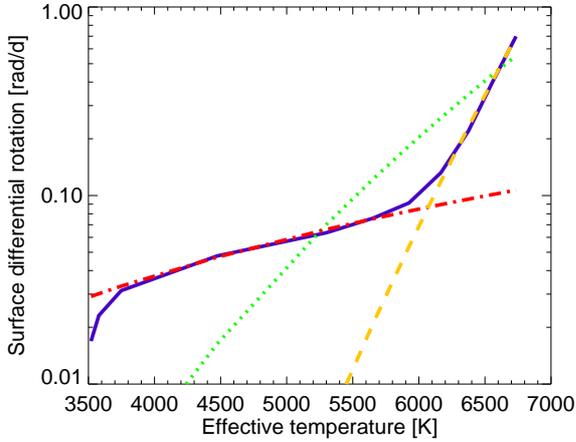}
\caption{ \label{domteff}
Surface differential rotation as a function of the effective temperature on the ZAMS for an equatorial rotation period of 2.5 d. The solid blue line denotes the result from the model, the dotted green line the power law by Collier Cameron (2008). The dash-dotted red line shows power law fits to the flat part of the blue line between 3500 and 6000 K, and the dashed yellow line a fit of the steep part for temperatures above 6000 K.
}
\end{figure}

%
%
%
We first compute the surface differential rotation for a sequence of ZAMS models with masses from 0.3 to 1.5 solar masses and an equatorial rotation period 2.5 d (ten times the solar value). The resulting relation between $\delta \Omega$ and the effective temperature is shown in   
Figure \ref{domteff}. We see a sharp increase of the surface differential rotation at temperatures below 3700 K, then a slower increase between 3800 and 5800 K, and a steep increase  at temperatures above 6000 K. The dash-dotted red and dashed yellow lines mark power law fits to parts of the solid blue curve.  
The dash-dotted red line denotes a power law of the form
\begin{equation} 
 \delta \Omega = 0.071   \left(\frac{T_{\rm eff} }{ 5500 \;{\rm K}}\right) ^ 2 {\rm rad/d}\end{equation}
which fits the slow increase between 3800 K and 5000 K remarkably well.
The dashed yellow line represents a power law with a much larger exponent, namely
\begin{equation}
  \delta \Omega = 0.012    \left(\frac{T_{\rm eff} }{ 5500 \;{\rm K}} \right) ^ {20} {\rm rad/d}
\end{equation}
that fits the region above 6000 K. While these power laws provide very good approximations of the model data in the respective regions for which they were computed, we did not find a single power law that reproduces the data reasonably well over the entire interval from 3800 K to 6700 K. The dotted green line shows the power law (\ref{barneslaw}), which clearly is not a good approximation to the solid blue line

\begin{figure}
\includegraphics[width=8cm]{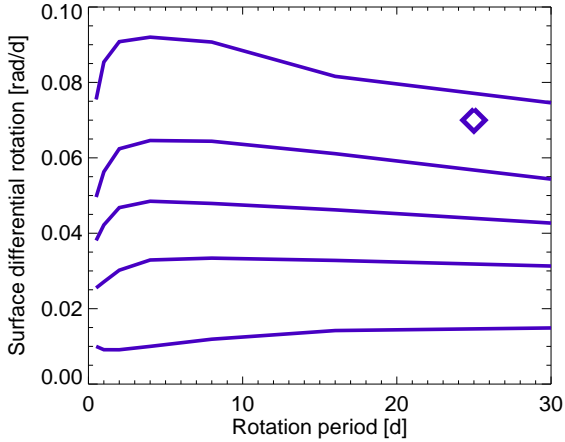}
\caption{ \label{domprot}
Surface differential rotation for ZAMS stars of 1.1, 0.9, 0.7, 0.5, and 0.3 solar masses (from top to bottom) vs.~the rotation period. The diamond indicates the present-day Sun.}
\end{figure}

Next we vary the rotation period for individual model stars to study how the surface differential rotation changes. Figure \ref{domprot} shows the result for ZAMS stars with masses from 0.3 to 1.1 solar masses. The spacing of the curves reflects the dependence on effective temperature shown in Fig.\ref{domteff}, with the top curve representing the 1.1 $M_\odot$ star and the bottom curve the 0.3 $M_\odot$ star. 
For each  star the variation with the rotation period is rather weak. 
For the stars with outer convection zones the differential rotation increases with the rotation period for very rapid rotation, goes through a maximum at periods between four andd eight days, and decreases for slow rotation.
For the fully convective 0.3 $M_\odot$ star we find the opposite pattern with a minimum between one and two days.
The differential rotation has been computed between the equator and $75^{\circ}$ latitude for this star instead of between the equator and the pole as for the other stars. This is because there is a spurious decrease of the rotation rate at the polar cap that may be caused by the inner boundary which still exists for technical reasons despite the fact that the star is fully convective. If the difference between equator and pole is taken the same way as for the more massive stars the differential rotation values are larger but the qualitative behavior is the same. 
For stars more massive than 1.1 $M_\odot$ and slow rotation (period longer than 8 d) we either did not find a solution or the rotation pattern was more complicated than the solar-type differential rotation found for fast rotation and lower masses.
%
%
%
%
\begin{figure}
  \includegraphics[width=8cm]{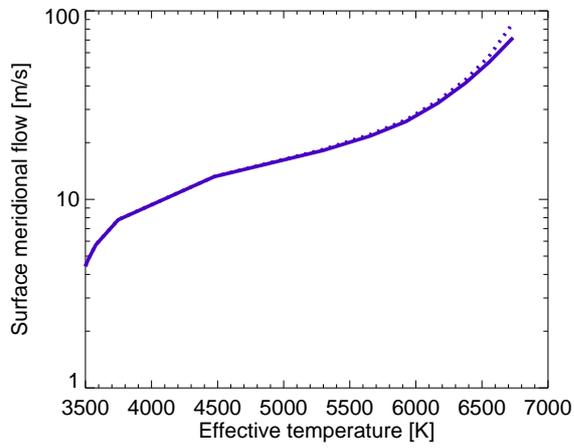}
  \caption{ \label{umedteff}
  Surface meridional flow speed vs.~effective temperature for ZAMS stars rotating with an equatorial period of 2.5 d. The solid line shows the full model, the dotted line the model with the baroclinic term dropped.
  }
  \end{figure}
\begin{figure}
  \includegraphics[width=8cm]{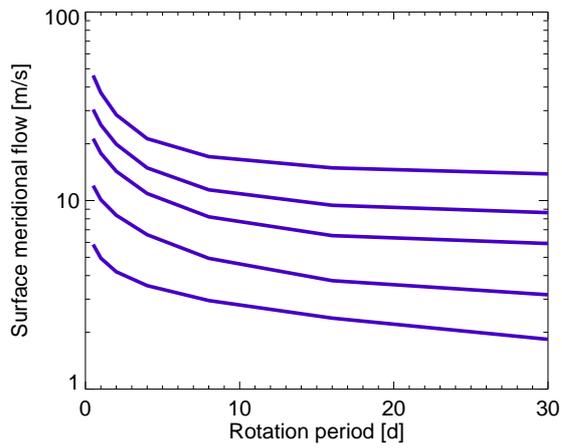}
  \caption{ \label{umedprot}
  Surface meridional flow speed vs.~equatorial rotation period for ZAMS star with 1.1, 0.9, 0.7, 0.5, and 0.3 solar masses (top to bottom).
  }
\end{figure}
The strong variation of the surface differential rotation with temperature hints at a similar relation between the meridional flow and the effective temperature. 
Fig.~\ref{umedteff} shows the maximum flow speed at the stellar surface versus the effective temperature for an equatorial rotation period of 2.5 d. 
As with the differential rotation, there is a pronounced increase with temperature which, however, is weaker. While the differential rotation increases by almost two orders of magnitude, the meridional flow increases by one only. Most notably, the sharp increase above 6000 K shown by the differential rotation is absent. The meridional flow increases faster at high temperatures, too, but the difference between the regimes below and above 6000 K is much smaller. The dotted line shows the flow speed for the same stars but with the baroclinic term dropped from the meridional flow equation. The results are almost identical with the full model except for high temperatures, where there is a noticeable (though still modest) increase. 

The relation between meridional flow and rotation is illustrated by Figure \ref{umedprot},  shows the maximum value of the surface meridional flow vs.~the rotation period for the same ZAMS models as Fig.~\ref{domprot}. 
As with the differential rotation, the curves represent masses from 1.1 (top) down to 0.3 (bottom) solar masses.   Unlike the differential rotation, the meridional flow does not show a maximum or minimum at a certain rotation period. Instead, the flow speed monotonously decreases with the rotation period. 
This is surprising at first sight but naturally results from the fact that the main driving force is the {\em square of} the angular velocity,
$ 2 \Omega {\partial \Omega}/{\partial z}$,
which decreases as $1/P_{\rm rot}$ for constant differential rotation. The monotonous decrease of the meridional flow is thus a consequence of the weak rotation dependence of the differential rotation. 

\begin{figure}
\includegraphics[width=8cm]{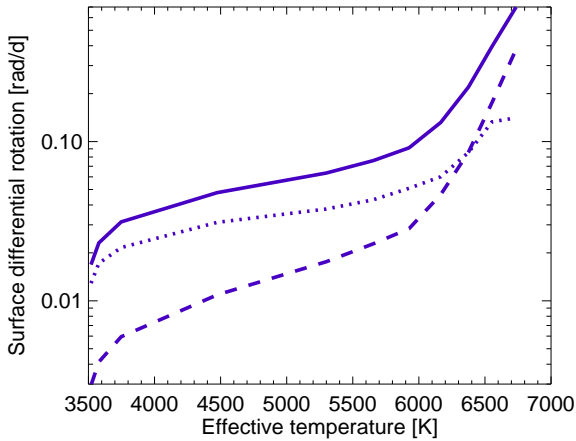}
\caption{ \label{contribs}
Differential rotation vs.~effective temperature for a rotation period of 2.5 d. Solid line: full model. Dotted line: baroclinic term cancelled. Dashed line: $\Lambda$ effect cancelled.}
\end{figure}

Both the $\Lambda$ effect and the baroclinic flow drive a solar-type differential rotation. To find out how much either effect contributes to the total shear we repeat the computations shown in Fig.~\ref{domteff} with the baroclinic term and the $\Lambda$ effect cancelled, respectively. The results are shown in Fig.~\ref{contribs}. For temperatures below 6000 K the $\Lambda$ effect clearly is the main generator of differential rotation. Above 6000 K the contribution from the baroclinic term increases fast and becomes dominant at around 6500 K.

\begin{figure}
   \begin{center}
     \includegraphics[width=8cm]{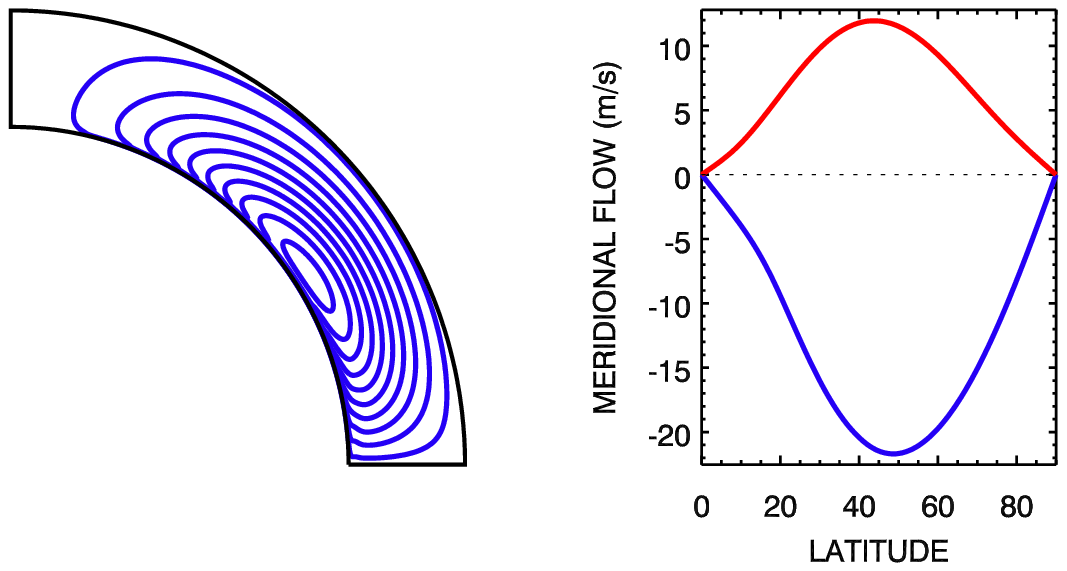} \\ 
     \includegraphics[width=8cm]{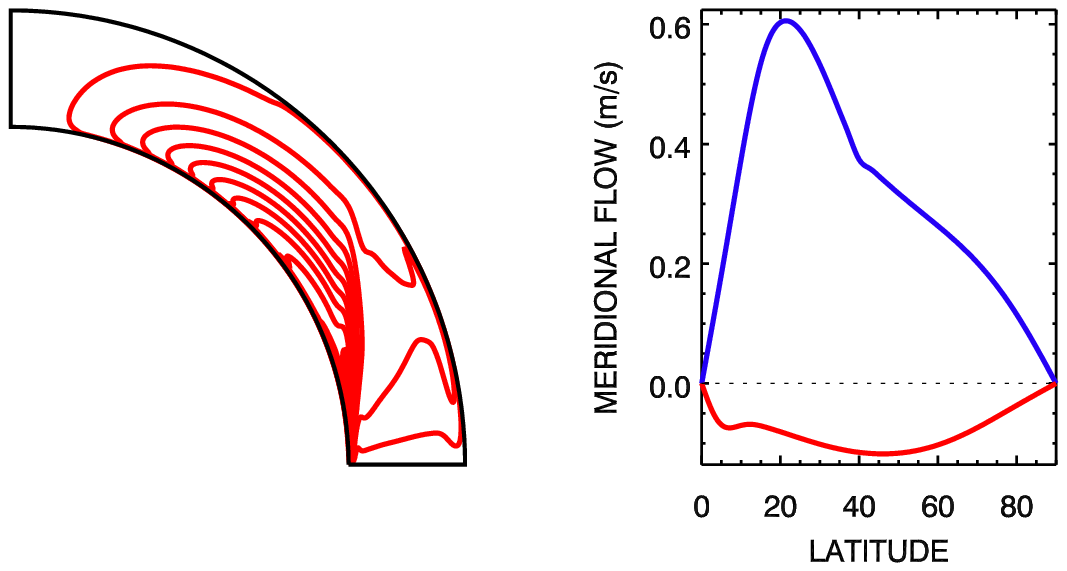}
   \end{center}
\caption{ \label{flow}
Meridional flow for a solar mass ZAMS model star rotating with a period of 2.5 d. Top: full model. Bottom: $\Lambda$ effect cancelled in the Reynolds stress. The left hand side diagrams show stream lines of the flow. Blue lines denote solar-type circulation, red lines anti-solar circulation. The diagrams on the right hand side show the meridional flow at the top (blue lines) and bottom (red lines) of the convection zone. Positive values refer to a gas motion towards the equator. 
}
\end{figure}

The Reynolds stress is also the main cause of the meridional flow. To illustrate this, we have computed the flow pattern for a solar-mass ZAMS star. Figure \ref{flow} shows the result for the full model (top panel) and with the $\Lambda$ effect cancelled (bottom panel). The flow from the full model resembles that found for the Sun, with a surface flow amplitude of about 20 m/s and the flow directed towards the pole. When the $\Lambda$ effect is dropped the meridional flow is much slower and the surface flow is directed towards the equator.  The flow is mostly confined to within the tangent cylinder around the lower boundary and the return flow is concentrated at the bottom of the convection zone.
%
%
\section{Conclusions}
We find a systematic dependence of the surface differential rotation on the effective temperature along the lower ZAMS. While the increase with temperature agrees qualitatively with the findings of Barnes et al.~(2005), we find a different quantitative behavior. The relation between $\delta \Omega$ and $T_{\rm eff}$ can not be fitted to a single power law. Instead we find a modest increase for low and a very sharp increase for higher temperatures. The large values we find above 6500 K  are in line with the observational findings for F stars by Reiners (2006). We do, however, not find the very small values that have been derived for some low mass stars (cf.~Collier Cameron 2008). 

Fully convective stars pose a challenge to our model which always includes a radiative core. This may be chosen as small as one percent of the stellar radius but its presence still causes a spurious drop of the rotation period at high latitudes. We therefore have not extended our study to temperatures below 3500 K. 

Our numerical experiments show that in the model the Reynolds stress is the main driver of differential rotation. Solar-type surface shear can also be caused by the baroclinic term alone, but in that case the meridional flow is very slow and "anti-solar," i.e.~directed towards the equator at the surface. Dropping the baroclinic term while keeping the $\Lambda$ effect, on the other hand, does not significantly change the meridional flow speed or pattern.  

The temperature dependence of the meridional flow is qualitatively similar to that of the differential rotation but does not show the steep increase at temperatures above 6000 K. This is accompanied by a growing discrepancy between the full model and the model with the baroclinic term switched off, as seen in Fig.~\ref{contribs}. This trend is seen in the meridional flow, too, where the maximum flow speed is 72 m/s for the full model but 86.3 m/s for the model without the baroclinic term, which becomes more important as the convection zone becomes shallow.   

The relative importance of the Reynolds stress and the baroclinicity can be seen in Fig.~\ref{wind}. While the bulk of the convection zone is in thermal wind equilibrium, the boundary layers are not. The large-scale meridional flow is thus driven by the boundary layers at the top and bottom of the convection zone. For our sequence of ZAMS models the baroclinic term turns out less important than for the Sun, where its cancellation leads to a much weaker differential rotation and isocontours of the rotation rate that are much more cylinder-shaped than the observed rotation pattern (K\"uker et al.~2011). This is obviously the consequence of the faster rotation, as the convection zone of a solar-mass ZAMS star does not differ fundamentally  from that of the present-day Sun. Repeating the experiment with a rotation period of 25 d indeed confirms this conclusion. Cancellation of the barocinic term leads to a reduction of the surface shear by as much as 80 \%.
     
The slowness of the meridional flow shown in the bottom part of Fig.~\ref{flow} is a consequence of the thermal wind equilibrium, which is a condition for the absence of meridional flows as it requires that  the force terms cancel out.
Any remaining meridional flow thus indicates the deviation from the thermal wind balance. Without the $\Lambda$ effect the deviation is indeed small but with the full Reynolds stress the situation is different as the much higher flow speed shows.
 
For the short rotation periods characteristic of young main sequence stars the baroclinic term is less effective than in the Sun, especially on the lower end of the ZAMS. For masses greater than one solar mass it has a distinct impact on the differential rotation but its influence on the meridional flow is small even then. This means that rapidly-rotating stars are farther away from thermal wind equilibrium than more slowly rotating stars like the Sun.

The strong increase of differential rotation with effective temperature lets us expect a corresponding increase of magnetic activity while the temperature dependence of the meridional flow would imply a decrease of the cycle period with effective temperature. As, however, the meridional flow is also strongly dependent on the rotation rate, the temperature dependence of the cycle time might not be easy to verify.
%
%

\end{document}